\documentclass[runningheads]{llncs}
\usepackage[colorlinks,
            linkcolor=red,
            anchorcolor=blue, 
            citecolor=blue,       
            ]{hyperref}

\usepackage{graphicx}
\usepackage{subfigure}
\usepackage{amsmath,amssymb}
\usepackage{multirow}
\usepackage{mathrsfs}
\usepackage[table,xcdraw]{xcolor}
\usepackage{float}
\usepackage{ragged2e}
\usepackage{marvosym}
\begin{document}
\title{Continuous longitudinal fetus brain atlas construction via implicit neural representation}
\titlerunning{Continuous atlas construction via implicit neural representation}
%
\author{Lixuan Chen \inst{1} \and Jiangjie Wu\inst{1} \and Qing Wu\inst{1} \and Hongjiang Wei\inst{2} \and Yuyao Zhang\textsuperscript{1,3(\Letter)}}
\authorrunning{L. Chen et al.}
\institute{School of Information Science and Technology, ShanghaiTech University, \\ Shanghai, China \and
School of Biomedical Engineering, Shanghai Jiao Tong University, Shanghai, China \and
Shanghai Engineering Research Center of Intelligent Vision and Imaging, ShanghaiTech University, Shanghai, China\\
\email{zhangyy8@shanghaitech.edu.cn}
}
\maketitle
\begin{abstract}
Longitudinal fetal brain atlas is a powerful tool for understanding and characterizing the complex process of fetus brain development. Existing fetus brain atlases are typically constructed by averaged brain images on discrete time points independently over time. Due to the differences in onto-genetic trends among samples at different time points, the resulting atlases suffer from temporal inconsistency, which may lead to estimating error of the brain developmental characteristic parameters along the timeline. To this end, we proposed a multi-stage deep-learning framework to tackle the time inconsistency issue as a 4D (3D brain volume + 1D age) image data denoising task. Using implicit neural representation, we construct a continuous and noise-free longitudinal fetus brain atlas as a function of the 4D spatial-temporal coordinate. Experimental results on two public fetal brain atlases (CRL and FBA-Chinese atlases) show that the proposed method can significantly improve the atlas temporal consistency while maintaining good fetus brain structure representation. In addition, the continuous longitudinal fetus brain atlases can also be extensively applied to generate finer 4D atlases in both spatial and temporal resolution.

\end{abstract}
\section{Introduction}
The development of the fetal brain is a complex and dynamic process~\cite{tilea2009cerebral,righini2006hippocampal}.
Abnormal development of the fetus brain may lead to long-term neurodevelopmental disorders and may even affect the quality of life in the perinatal and later childhood. Longitudinal fetal brain atlas is an important tool to boost the understanding of fetus brain development and provide a statistical standard of fetus brain structure at different gestational ages. There have been few longitudinal fetus brain atlases~\cite{CRL,FBA,habas2010spatiotemporalatlas,serag2012atlas}, which constructed the averaging templates at discrete time points independently over time or simply added smoothing kernels on the age window. Due to the differences in onto-genetic trends among samples at different time points, the effect of noise along the developmental timeline is one of the most critical challenges for longitudinal atlas construction. Besides, limited by the image reconstruction quality of individual fetus brain, atlas quality may also suffer from reconstruction artifacts. Such issues will result in a certain precision error when quantifying the developmental characteristic parameters at each time point.\\
\indent To address the temporal inconsistency issue in longitudinal atlas construction, Zhang et al.~\cite{zhang2016consistent} proposed a 4D infant brain atlas construction method via introducing a temporal consistency term in the atlas sparse reconstruction. By incorporating the longitudinal constraint on a learning-based registration method, Chen et al.~\cite{miccaicerebellumatlases} proposed an age-conditional learning framework to construct 4D infant cerebellum atlas. Similarly, Zhao, F, et al.~\cite{miccaisurfaceatlas} proposed a similar temporal constraint on an unsupervised learning-based surface atlas construction. 
However, most of these mentioned works significantly depend on the specific data collection process. Specifically, these methods usually need a sequence of scans from the same subjects within the age range of interest, which are typically expensive and super challenging to acquire.\\
\indent From another perspective, this problem could be modeled as a single four-dimensional (4D: 3D brain volume + 1D age) image data denoising problem that emphasizing to reduce noise along the timeline. This is because, the 3D image noise is largely reduced during the averaging template generation process, where the temporal consistency is normally not properly considered. The single image self-supervised denoising problem has been well-studied over the past several years. Ulyanov et al.~\cite{ulyanov2018dip} introduced deep image prior (DIP) to solve denoising problems by fitting the weights of an over-parameterized deep convolutional network to a single image, together with strong regularization by early stopping of the optimization. However, the hyper-parameters of early stopping are hard to tune, and the high-frequency content reduced by early stopping could be both noise and image details. In~\cite{lehtinen2018noise2noise}, Noise2Noise method proposes a statistically more meaningful manner to reduce only zero-mean image noise by learning the differences between two noisy observations of the same object. For avoiding using noisy image pairs as in~\cite{lehtinen2018noise2noise}, several single image denoising methods are designed to build specific blind-spot network structures to decrease image noise~\cite{huang2021neighbor2neighbor,krull2019noise2void}.\\
\indent In this work, we propose a multi-stage learning framework to train and refine a continuous longitudinal fetus brain atlas that is continuous in both 3D coordinate and timeline. Overall, we iteratively refine the existing longitudinal fetus brain atlas with the temporal inconsistency problem via a 4D single image denoising task. Specifically, 1) As brain growth is highly continuous and follows certain trajectories, which could be fitted by a continuous function. 
We then use spatially encoded multi-layer perceptron network (MLP)~\cite{IREM} to implicitly represent the longitudinal atlas as a 4D (3D coordinate and time t) continuous image function; 2) However, the above method has the same issue as DIP~\cite{ulyanov2018dip}. The network has sufficient capacity to overfit image noise without early stopping.
Thus, we divide the original longitudinally atlas into two different groups according to time points and approximate the 4D images in each of the two separated groups using two continuous functions. Since the two sets of atlases represent the same brain developing trajectory, the learned functions are theoretically equivalent. 
However, since the two networks overfit different image noise, two different functions are actually learned. 
By encouraging the atlases inferred by these two continuous functions at arbitrary new time point to be the same, we induce atlases with reduced noise along gestational age, i.e., better temporally-consistency; 3) By averaging the two final continuous functions, we construct the continuous longitudinal fetus brain atlas function. To the best of our knowledge, it is the first time to tackle the temporal inconsistency as a 4D image denoising problem and improve the consistency of existing atlases. Both qualitative and quantitative results demonstrate that comparing to the original atlas, the refined atlas achieves better time consistency while maintaining a good representation of fetus brain structure.
\begin{figure}[t]
\includegraphics[width=\textwidth]{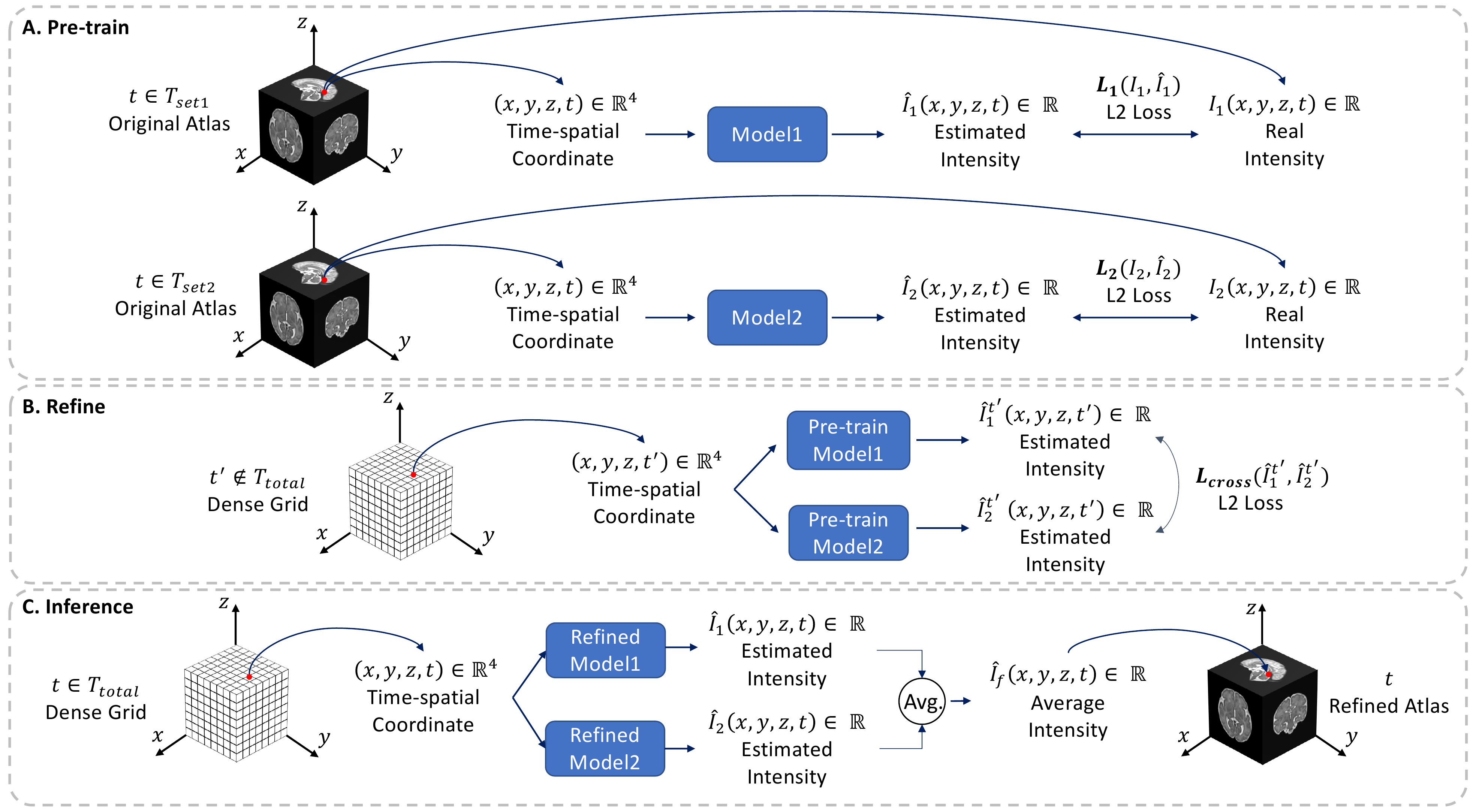}
\caption{An overview of the proposed continuous longitudinal fetus brain atlas construction framework.} \label{fig1}
\end{figure}

\section{Method}
An overview of our framework is depicted in Fig.~\ref{fig1}, which is presented in three stages: a) Pre-train stage; b) Refine stage; c) Inference stage.\\
\indent In this paper, we present a very simple yet effective method to train and refine a continuous 4D fetus brain atlas.  Firstly, the original atlas set that includes $N$ different time points $T_{total}=\left\{t_{i}\right\}_{i=1}^{N}$ are divided into two separated subgroups $T_{set1}=\left\{t_{2n-1}\right\}$ and $T_{set2}=\left\{t_{2n}\right\}$, where $n=1,...,N/2$ (as Fig.~\ref{fig1}(A)). Then images in each of the group are approximated using a 4D continuous function $\hat{I}_{1}=f_{\theta_1}(x,y,z,t),t\in T_{set1}$ and $\hat{I}_{2}=f_{\theta_2}(x,y,z,t),t\in T_{set2}$.  Secondly, series of novel time dependent brain images 
at time points $t'\notin T_{total}$ are generated as $\hat{I}_{1}^{t'} = f_{\theta_1}(x,y,z,t')$ and $\hat{I}_{2}^{t'} = f_{\theta_2}(x,y,z,t')$ (Fig.~\ref{fig1}(B)). A denoising network is then trained by input image $\hat{I}_{1}^{t'}$ and labeled by image $\hat{I}_{2}^{t'}$ at the same time series of $t'$. The two stream refinement strategy satisfying the image denoising requirement as in~\cite{lehtinen2018noise2noise} that paired pixels of paired images with independent noise and equivalent image content, and avoiding from using a second noisy image observation. Finally, by averaging the two continuous fetus brain function we learnt and refined, we construct the final temporally-continuous 4D fetus brain atlas function as $f_{\theta_{mean}}(x,y,z,t)= 1/2 f_{\theta_1}(x,y,z,t) + 1/2 f_{\theta_2}(x,y,z,t)$. By passing the series of time points as that from the original atlas series into the continuous atlas function $\hat{I}_{f} = f_{\theta_{mean}} (x,y,z,t), t\in T_{total}$, we reconstruct a longitudinally-consistent 4D series of fetus brain atlases (Fig.~\ref{fig1}(C)).

\begin{figure}[t]
\centering
\includegraphics[width=0.9\textwidth]{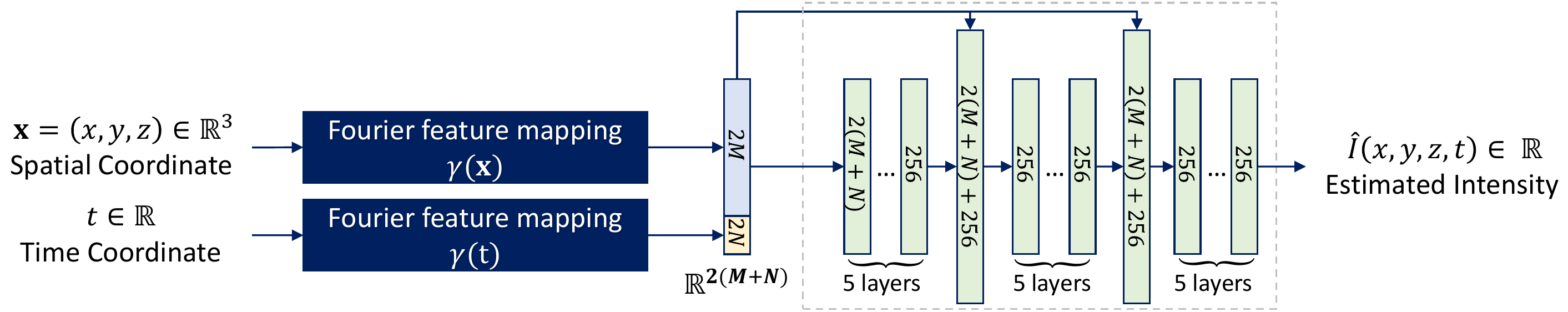}
\caption{Architecture of our model} \label{fig2}
\end{figure}

\subsection{Pre-train stage}
As demonstrated in Fig.~\ref{fig1}(A), we first divide $T_{total}$ into $T_{set1}$ and $T_{set2}$. Then, we use two spatial encoded MLPs to approximate the 4D image per set as a continuous function. Specifically, we feed the voxel coordinate and time $(x,y,z,t)$ into our model $f_\theta$ to compute the predicted voxel intensity $\hat{I}(x,y,z,t)$. We donate $\hat{I}$ as a explicit 4D image and $\hat{I}(x,y,z,t)$ as the intensity at location $(x,y,z)$ and time $t$. We learn the parameters $\theta$ by minimizing the mean square error (MSE) loss function between the predicted voxel intensity and the real observed voxel intensity at current time point for a mini-batch of size, $P$. The loss function $\mathcal{L}$ is denoted as: 
\begin{equation}
\mathcal{L}(\theta)=\frac{1}{|P|}\sum\limits_{(x,y,z,t)\in P}||I(x,y,z,t)-\hat{I}(x,y,z,t)||^2
\end{equation}
\textbf{Architecture of our Model.} The architecture and training strategy is exactly same for Model1 and Model2 in Fig.~\ref{fig1}(A). As illustrated in Fig.~\ref{fig2}, our model consists of a encoding section and a MLP network, taking the 4D time-spatial coordinate input and outputting the corresponding intensity. For the encoding section, Mildenhall et al.~\cite{tancik2020fourier} recently proposed Fourier feature mapping to overcome the spectral bias that the standard MLPs are biased towards learning lower frequency functions~\cite{rahaman2019spectral}. 
In our model, we perform Fourier feature mapping to respectively map the 3D voxel coordinates and the time $t$ to the higher dimensional space $\mathbb{R}^{2L}$($2L>x$, $x=1$ or 3) before passing them to the standard MLP network. Let $\gamma(\cdot)$ denotes Fourier feature mapping from the space $\mathbb{R}^x$ to $\mathbb{R}^{2L}$ and it is calculated by $\gamma(\mathcal{\textbf{v}})=[\cos(2\pi \textbf{Bv}),\sin(2\pi \textbf{Bv})]^T$
where $\textbf{v}\in\mathbb{R}^x$ and each element in $\textbf{B}\in\mathbb{R}^{L\times x}$ is independently sampled from standard normal distribution $\mathcal{N}(0,1)$. For the MLP network, the network has eighteen fully-connected layers with two skip connections~\cite{he2016skipconnectiona,he2016skipconnectionb} that concatenate the input of the fully-connected network to the $6^{th}$ and $12^{th}$ layer’s activation. Each fully-connected layer is followed by a batch normalization~\cite{ioffe2015batch} layer and a ReLU~\cite{nair2010ReLU} activation.

\begin{figure}[t]
\includegraphics[width=\textwidth]{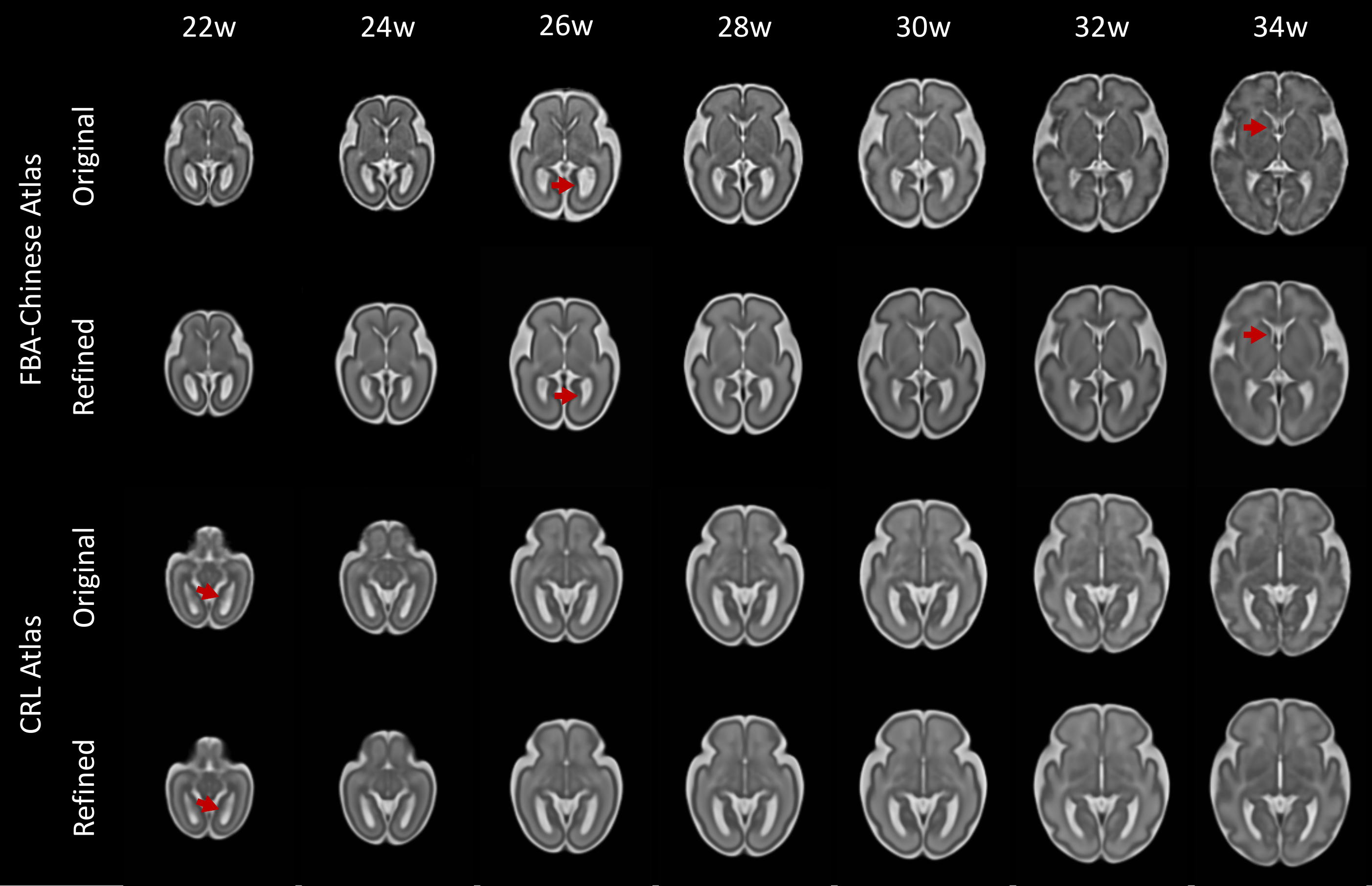}
\caption{Visual comparison overview of original and refined CRL and FBA-Chinese atlases. Note that although limited time points are shown, our framework provides temporally-continuous 4D atlases at arbitrary time points. } \label{fig3}
\end{figure}

\subsection{Refine stage}
The “refine” strategy is illustrated in Fig.~\ref{fig2}(B). After the processing in Sec. 2.1, the two pre-trained models have already overfitted to the two subsets of the original longitudinally atlas with noise. Ideally, the two pre-trained models should approximate the same continuous function since the two 4D sub-images are sampled from the same 4D singe image (original longitudinal atlases) and represent the same brain developing trajectory. However, due to the models in the “pre-trained” stage overfit to different image noise, the two networks are slightly different in fact. Inspired by Noise2Noise~\cite{lehtinen2018noise2noise}, we propose a “refine” stage to refine the pre-trained models and find the final continuous function, which could represent the continuous time-consistent 4D atlas.\\
\indent We feed the image coordinate grid $(x,y,z)$ with arbitrary new time points $t'\notin T_{total}$ to these two pre-trained models to generate two series of time dependent brain images $\hat{I}_{1}^{t'} = f_{\theta_1}(x,y,z,t')$ and $\hat{I}_{2}^{t'} = f_{\theta_2}(x,y,z,t')$. Because those two models tend to approximate the same brain developing trajectory, then we suppose the predicted voxel intensity in $\hat{I}_{1}^{t'}$ and $\hat{I}_{2}^{t'}$ should be identical. So we further iteratively update the parameters of the two pre-trained functions by minimizing the mean square error(MSE) loss function between the two predicted brain images at new time points. The formulation of loss function is similar to Eq.(1). Besides, we design an updated cut-off condition $\mathcal{L}_{total}$:
\begin{equation}
\mathcal{L}_{total}=\lambda \cdot \mathcal{L}_1+\lambda \cdot \mathcal{L}_2+\mathcal{L}_{cross}
\end{equation}
where $\mathcal{L}_1$ and $\mathcal{L}_2$ are the image fidelity loss between MLP functions and the corresponding real observations at $T_{set1}$ and $T_{set2}$. While $\mathcal{L}_{cross}$ is the MSE loss between two predicted images $\hat{I}_{1}^{t'}$ and $\hat{I}_{2}^{t'}$ from two pre-trained models. $\lambda$ is a hyper-parameter used to adjust the proportion of each loss. When $\mathcal{L}_{total}$ is minimal, the model we get is optimal.
\subsection{Inference stage}
The sketch map of a longitudinally-consistent 4D fetus brain atlas reconstruction by the final refined model is shown in Fig.~\ref{fig2}(C). Once the training is completed, we average the two continuous functions and construct the final continuous 4D fetus brain atlas function as $f_{\theta_{mean}}(x,y,z,t)= 1/2 f_{\theta_1}(x,y,z,t) + 1/2 f_{\theta_2}(x,y,z,t)$. Then given the age attribute $t$ in $T_{total}$, we pass the voxel coordinate $(x,y,z)$ with $t$ into our final model to reconstruct a longitudinally-consistent 4D series of fetus brain atlases $\hat{I}_{f} = f_{\theta_{mean}} (x,y,z,t), t\in T_{total}$. 

\section{Experiments}
\subsection{Setup}
\textbf{Dataset.} We evaluated the proposed framework on two existing public longitudinal fetus brain atlases: CRL atlas~\cite{CRL} and FBA-Chinese atlas~\cite{FBA}. CRL is constructed from MRI of 81 normal Caucasian fetuses scanned between 21 and 38 weeks of gestation. FBA-Chinese atlas is created from 115 normal Chinese fetal brains between 22 and 34 weeks of gestation. 

\noindent\textbf{Implementation Details.} For CRL atlas, we set $T_{set1}$=$\{$21, 23, 25, 27, 29, 31, 33, 35, 37, 38 week$\}$, $T_{set2}$=$\{$21, 22, 24, 26, 28, 30, 32, 34, 36, 38 week$\}$ and arbitrary new time set= $\{t+0.5\}_{t=21}^{37}$.
For FBA-Chinese atlas, we set $T_{set1}$=$\{$22, 24, 26, 28, 30, 32, 34, 35 week$\}$, $T_{set2}$=$\{$22, 23, 25, 27, 29, 31, 33, 35 week$\}$ and arbitrary new time set= $\{t+0.5\}_{t=22}^{34}$.\\
\indent In “pre-train” stage, the Fourier feature mapping dimension $2L$ is set as 256 for 3D coordinate and 64 for time. Besides, our models are trained with a batch size of 25000 using an Adam~\cite{kingma2014adam} optimizer with $\beta = (0.9,0.999)$. The learning rate starts from $10^{-4}$ and decays by factor 0.5 every 100 epochs. In “refine” stage, we set $\lambda$ as 0.1 in cut-off condition $L_{total}$.

\begin{figure}[t]
\centering
\subfigure[Cortical plate]{
\begin{minipage}[t]{0.32\linewidth}
\centering
\includegraphics[width=\textwidth]{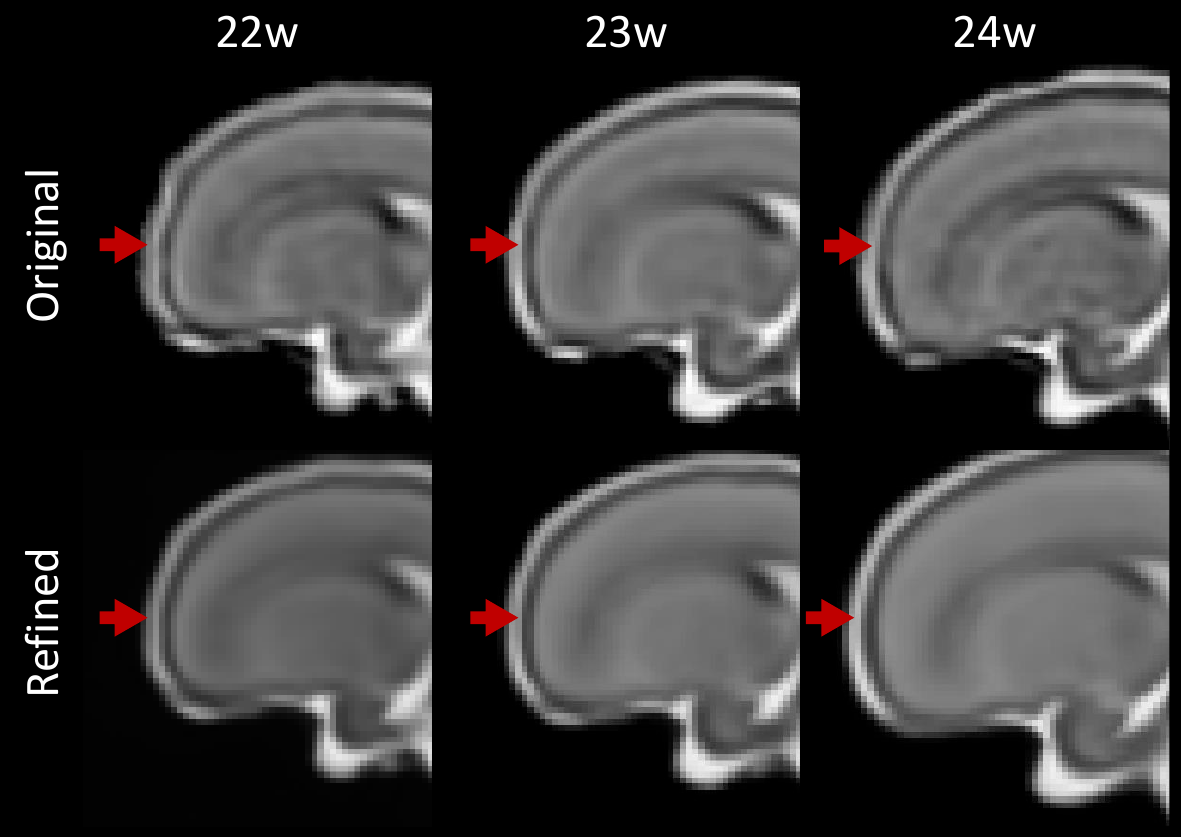}
\end{minipage}
}%
\subfigure[Residual skull]{
\begin{minipage}[t]{0.32\linewidth}
\centering
\includegraphics[width=\textwidth]{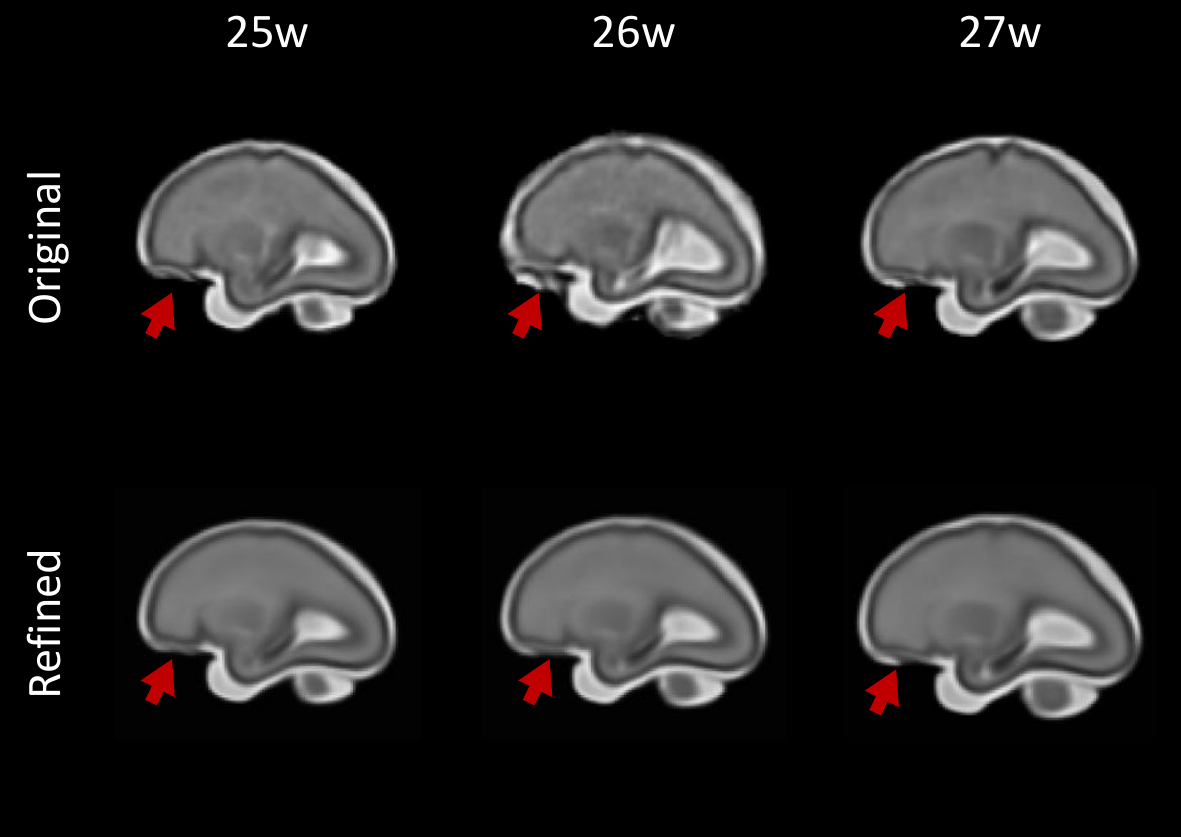}
\end{minipage}
}%
\subfigure[Cerebellar boundary]{
\begin{minipage}[t]{0.32\linewidth}
\centering
\includegraphics[width=\textwidth]{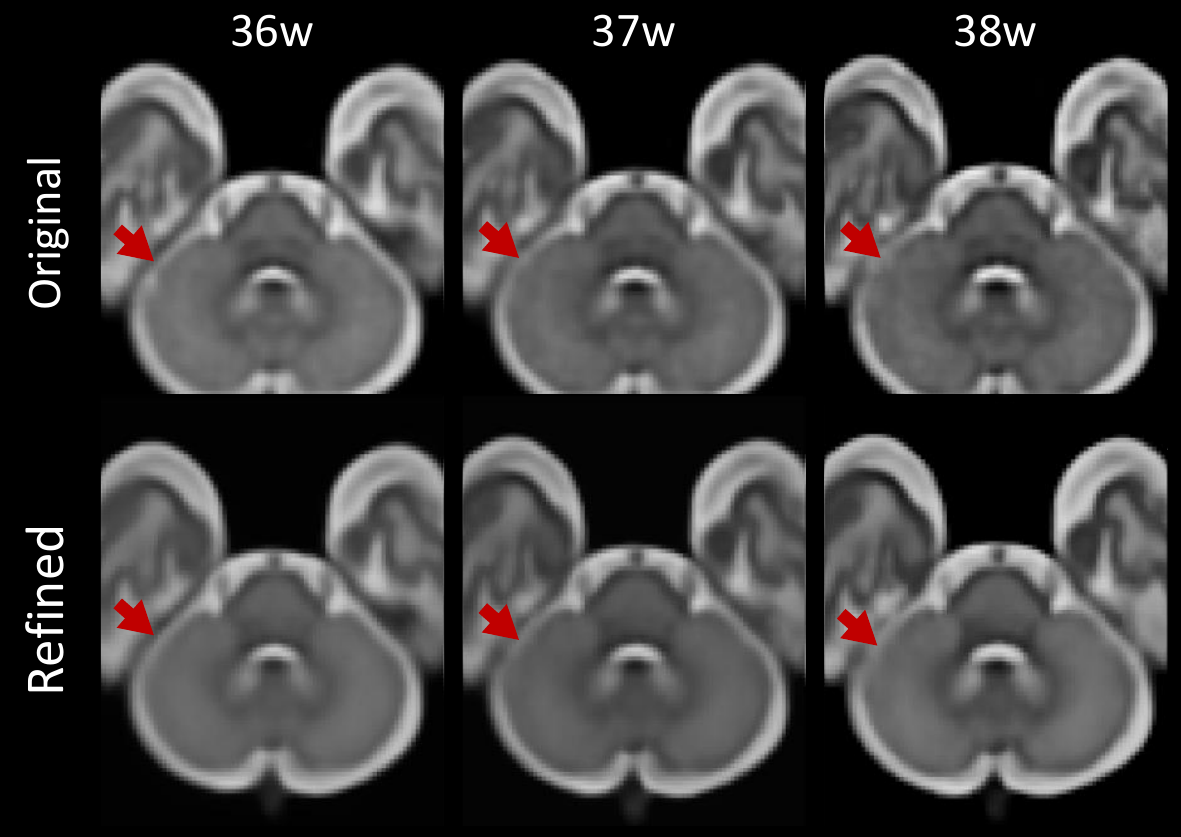}
\end{minipage}%
}%
\centering
\caption{Supplementary visual comparison in other views. (a) Comparison of cortical plate in CRL atlas. (b) Comparison of residual skull in FBA-Chinese atlas. (c) Comparison of cerebellar boundary in FBA-Chinese atlas.}
\label{fig4}
\end{figure}

\subsection{Results}
\textbf{Visual Comparisons.} Fig.~\ref{fig3} provides typical axial examples of the CRL and FBA-Chinese atlases refined by proposed framework.\\
\indent Generally, the proposed method illustrates clear denoising performance on both 3D image volume and 1D time line. For example, the image contrast in the basal ganglia regions is improved in the reconstructed brains. Besides, as arrow high-lighted, for the original \textbf{FBA-Chinese} atlas, the lateral ventricle of 26w atlas is significantly larger than that of 25w and 27w. The proposed model corrected this brain structure noise. Similarly, the vascular detail at 34w is refined. Figure 4 provides the results of other views and more detailed refinement in brain anatomical structures. Fig.~\ref{fig4}(a) shows that the cortical plate of original atlases has artifacts caused by motion, and the atlases obtained through our framework have a more consistent cortical plate. Fig.~\ref{fig4}(b) shows that our framework can effectively improve the inconsistency problem caused by incomplete skull stripping during atlas construction. For \textbf{CRL} atlas, the image contrast is improved at age 22w and ventricle structure is also refined. Besides, as can be seen from Fig.~\ref{fig4}(c), the refined atlases obtain a more apparent cerebellar boundary.

\noindent \textbf{Quantitative Comparision.} 
We used three evaluation matrices: Entropy Focus Criterion (EFC), DICE coefficients and Time Consistency(TC) factor to evaluate our longitudinal fetus brain atlases. The results show that the atlases obtained by our method have better time consistency while ensuring good sharpness and representation. Detailed results are summarized in Table~\ref{tab1}.

To evaluate the sharpness of our proposed atlases, we use Entropy Focus Criterion (EFC)~\cite{atkinson1997automatic,sadri2020mrqy} in MRQy~\cite{sadri2020mrqy}. The entropy focus criterion of the entire atlas image is defined as $EFC=\frac{1}{M}\sum_{m=1}^M(|S*\frac{1}{\sqrt{S}}\ln{\frac{1}{\sqrt{S}}}|)^{-1}*(- \sum_{s=1}^S\frac{x_s}{x_{max}}\ln{\left[\frac{x_s}{x_{max}}\right]})$
where $S$ and $M$ is the number of voxels and slices, $m$ is the $m^{th}$ slice, and $x_{max}$ is defined by $x_{max}=\sqrt{\sum_{s=1}^S x_s^2}$ with signal intensity $x_s$. The images that lack sharp distinctions between brain regions would have high entropy values~\cite{liu2015low}. Table~\ref{tab1} shows that the refined atlases have similar or even lower values than original atlases, indicating our refined atlases have a relatively comparable image quality with original atlases in terms of sharpness.


To evaluate the representativeness of the refined fetal brain atlas, we performed the atlas-based segmentation method~\cite{FBA} with Dice similarity coefficient to evaluate. We use the Fetal Tissue Annotation and Segmentation Dataset (FeTA)~\cite{payette2021FeTA}, which contains 50 manually segmented fetal brains across 20 to 33 weeks with 7 different regions-of-interest (ROI). Specifically, these subject images with ROI label were firstly registered to the refined atlas and the original atlas, respectively. Then DICE coefficients between different subjects in same atlas space are calculated and compared. A higher DICE value indicates a higher accuracy of the atlas for guiding brain anatomical normalization. Meanwhile, a higher DICE value can also indicate that the atlas has better consistency because the atlas with better consistency would lead to better registration result. Table~\ref{tab1} shows that the refined atlases have higher DICE than original atlases

\begin{table}[t]
\caption{Quantitative comparison on CRL and FBA-Chinese atlases for sharpness(EFC), representation(DICE) and time consistency(TC). The best results are indicated in red.}\label{tab1}
\resizebox{\textwidth}{15mm}{\begin{tabular}{ccc|cccccccccccccccccc}
\cline{4-21}
                                                     &                                                      &      & \multicolumn{1}{c|}{\textbf{21}}                        & \multicolumn{1}{c|}{\textbf{22}} & \multicolumn{1}{c|}{\textbf{23}} & \multicolumn{1}{c|}{\textbf{24}} & \multicolumn{1}{c|}{\textbf{25}} & \multicolumn{1}{c|}{\textbf{26}} & \multicolumn{1}{c|}{\textbf{27}} & \multicolumn{1}{c|}{\textbf{28}} & \multicolumn{1}{c|}{\textbf{29}} & \multicolumn{1}{c|}{\textbf{30}} & \multicolumn{1}{c|}{\textbf{31}} & \multicolumn{1}{c|}{\textbf{32}} & \multicolumn{1}{c|}{\textbf{33}} & \multicolumn{1}{c|}{\textbf{34}}                  & \multicolumn{1}{c|}{\textbf{35}}                        & \multicolumn{1}{c|}{\textbf{36}}                        & \multicolumn{1}{c|}{\textbf{37}}                        & \textbf{38}                        \\ \hline
\multicolumn{1}{|c|}{}                               & \multicolumn{1}{c|}{}                                & ori. & {\color[HTML]{FE0000} 0.2018}                           & {\color[HTML]{FE0000} 0.2170}    & {\color[HTML]{FE0000} 0.2218}    & {\color[HTML]{FE0000} 0.2429}    & 0.2722                           & {\color[HTML]{FE0000} 0.2907}    & {\color[HTML]{FE0000} 0.2888}    & 0.3043                           & 0.3166                           & 0.3383                           & {\color[HTML]{FE0000} 0.3405}    & {\color[HTML]{FE0000} 0.3747}    & {\color[HTML]{FE0000} 0.3773}    & {\color[HTML]{FE0000} 0.3941}                     & 0.4095                                                  & {\color[HTML]{FE0000} 0.4107}                           & {\color[HTML]{FE0000} 0.4107}                           & 0.4114                             \\
\multicolumn{1}{|c|}{}                               & \multicolumn{1}{c|}{\multirow{-2}{*}{\textbf{EFC}}}  & ref. & 0.2049                                                  & 0.2181                           & 0.2294                           & 0.2435                           & {\color[HTML]{FE0000} 0.2604}    & 0.2932                           & 0.2929                           & {\color[HTML]{FE0000} 0.2899}    & {\color[HTML]{FE0000} 0.2912}    & {\color[HTML]{FE0000} 0.3241}    & 0.3548                           & 0.3771                           & 0.3827                           & 0.3989                                            & {\color[HTML]{FE0000} 0.4071}                           & 0.4140                                                  & 0.4113                                                  & {\color[HTML]{FE0000} 0.3646}      \\ \cline{2-21} 
\multicolumn{1}{|c|}{}                               & \multicolumn{1}{c|}{}                                & ori. & 37.27                                                   & 34.50                            & 42.22                            & 41.74                            & 30.05                            & 49.76                            & 49.02                            & 48.11                            & 44.86                            & 44.09                            & 60.01                            & 66.68                            & 54.43                            & \multicolumn{1}{c|}{63.81}                        & \multicolumn{1}{c|}{}                                   & \multicolumn{1}{c|}{}                                   & \multicolumn{1}{c|}{}                                   &                                    \\
\multicolumn{1}{|c|}{}                               & \multicolumn{1}{c|}{\multirow{-2}{*}{\textbf{DICE}}} & ref. & {\color[HTML]{FE0000} 41.66}                            & {\color[HTML]{FE0000} 37.34}     & {\color[HTML]{FE0000} 44.99}     & {\color[HTML]{FE0000} 43.39}     & {\color[HTML]{FE0000} 40.39}     & {\color[HTML]{FE0000} 52.05}     & {\color[HTML]{FE0000} 51.76}     & {\color[HTML]{FE0000} 49.17}     & {\color[HTML]{FE0000} 46.69}     & {\color[HTML]{FE0000} 47.51}     & {\color[HTML]{FE0000} 63.55}     & {\color[HTML]{FE0000} 71.00}     & {\color[HTML]{FE0000} 57.03}     & \multicolumn{1}{c|}{{\color[HTML]{FE0000} 73.52}} & \multicolumn{1}{c|}{\multirow{-2}{*}{\textbackslash{}}} & \multicolumn{1}{c|}{\multirow{-2}{*}{\textbackslash{}}} & \multicolumn{1}{c|}{\multirow{-2}{*}{\textbackslash{}}} & \multirow{-2}{*}{\textbackslash{}} \\ \cline{2-21} 
\multicolumn{1}{|c|}{}                               & \multicolumn{1}{c|}{}                                & ori. & 88.85                                                   & 91.15                            & 92.13                            & 92.17                            & 93.09                            & 93.80                            & 94.76                            & 94.92                            & 94.88                            & {\color[HTML]{FE0000} 94.67}     & 93.57                            & 93.41                            & 93.41                            & 93.34                                             & 93.15                                                   & 93.91                                                   & 93.54                                                   & 92.99                              \\
\multicolumn{1}{|c|}{\multirow{-6}{*}{\textbf{CRL}}} & \multicolumn{1}{c|}{\multirow{-2}{*}{\textbf{TC}}}   & ref. & {\color[HTML]{FE0000} 89.31}                            & {\color[HTML]{FE0000} 91.92}     & {\color[HTML]{FE0000} 92.94}     & {\color[HTML]{FE0000} 92.75}     & {\color[HTML]{FE0000} 93.32}     & {\color[HTML]{FE0000} 94.31}     & {\color[HTML]{FE0000} 94.79}     & {\color[HTML]{FE0000} 95.17}     & {\color[HTML]{FE0000} 94.98}     & {\color[HTML]{FE0000} 94.67}     & {\color[HTML]{FE0000} 93.82}     & {\color[HTML]{FE0000} 93.66}     & {\color[HTML]{FE0000} 93.73}     & {\color[HTML]{FE0000} 93.69}                      & {\color[HTML]{FE0000} 93.40}                            & {\color[HTML]{FE0000} 94.20}                            & {\color[HTML]{FE0000} 93.82}                            & {\color[HTML]{FE0000} 93.57}       \\ \hline
\multicolumn{1}{|c|}{}                               & \multicolumn{1}{c|}{}                                & ori. & \multicolumn{1}{c|}{}                                   & 0.2145                           & 0.2158                           & 0.2323                           & 0.2652                           & 0.2794                           & 0.2910                           & 0.3089                           & 0.3179                           & 0.3405                           & 0.3472                           & 0.3640                           & {\color[HTML]{FE0000} 0.3619}    & {\color[HTML]{FE0000} 0.3764}                     & \multicolumn{1}{c|}{0.4027}                             & \multicolumn{1}{c|}{}                                   & \multicolumn{1}{c|}{}                                   &                                    \\
\multicolumn{1}{|c|}{}                               & \multicolumn{1}{c|}{\multirow{-2}{*}{\textbf{EFC}}}  & ref. & \multicolumn{1}{c|}{}                                   & {\color[HTML]{FE0000} 0.2138}    & {\color[HTML]{FE0000} 0.2153}    & {\color[HTML]{FE0000} 0.2079}    & {\color[HTML]{FE0000} 0.2312}    & {\color[HTML]{FE0000} 0.2516}    & {\color[HTML]{FE0000} 0.2685}    & {\color[HTML]{FE0000} 0.2759}    & {\color[HTML]{FE0000} 0.2927}    & {\color[HTML]{FE0000} 0.3238}    & {\color[HTML]{FE0000} 0.3459}    & {\color[HTML]{FE0000} 0.3552}    & 0.3627                           & 0.3797                                            & \multicolumn{1}{c|}{{\color[HTML]{FE0000} 0.3761}}      & \multicolumn{1}{c|}{}                                   & \multicolumn{1}{c|}{}                                   &                                    \\ \cline{2-3} \cline{5-18}
\multicolumn{1}{|c|}{}                               & \multicolumn{1}{c|}{}                                & ori. & \multicolumn{1}{c|}{}                                   & 35.34                            & 42.85                            & 40.67                            & 31.59                            & 52.29                            & 48.57                            & 47.35                            & 44.49                            & 46.86                            & 62.27                            & 67.84                            & 53.67                            & \multicolumn{1}{c|}{64.62}                        & \multicolumn{1}{c|}{}                                   & \multicolumn{1}{c|}{}                                   & \multicolumn{1}{c|}{}                                   &                                    \\
\multicolumn{1}{|c|}{}                               & \multicolumn{1}{c|}{\multirow{-2}{*}{\textbf{DICE}}} & ref. & \multicolumn{1}{c|}{}                                   & {\color[HTML]{FE0000} 36.71}     & {\color[HTML]{FE0000} 44.71}     & {\color[HTML]{FE0000} 42.17}     & {\color[HTML]{FE0000} 32.72}     & {\color[HTML]{FE0000} 52.52}     & {\color[HTML]{FE0000} 52.34}     & {\color[HTML]{FE0000} 51.44}     & {\color[HTML]{FE0000} 45.88}     & {\color[HTML]{FE0000} 47.51}     & {\color[HTML]{FE0000} 62.85}     & {\color[HTML]{FE0000} 73.01}     & {\color[HTML]{FE0000} 58.54}     & \multicolumn{1}{c|}{{\color[HTML]{FE0000} 75.17}} & \multicolumn{1}{c|}{\multirow{-2}{*}{\textbackslash{}}} & \multicolumn{1}{c|}{}                                   & \multicolumn{1}{c|}{}                                   &                                    \\ \cline{2-3} \cline{5-18}
\multicolumn{1}{|c|}{}                               & \multicolumn{1}{c|}{}                                & ori. & \multicolumn{1}{c|}{}                                   & 91.27                            & 92.23                            & {\color[HTML]{FE0000} 92.09}     & 91.26                            & 91.31                            & 92.74                            & 92.96                            & 91.21                            & 92.44                            & 90.79                            & 90.96                            & 87.16                            & 86.59                                             & \multicolumn{1}{c|}{83.25}                              & \multicolumn{1}{c|}{}                                   & \multicolumn{1}{c|}{}                                   &                                    \\
\multicolumn{1}{|c|}{\multirow{-6}{*}{\textbf{FBA}}} & \multicolumn{1}{c|}{\multirow{-2}{*}{\textbf{TC}}}   & ref. & \multicolumn{1}{c|}{\multirow{-6}{*}{\textbackslash{}}} & {\color[HTML]{FE0000} 91.47}     & {\color[HTML]{FE0000} 93.08}     & 91.99                            & {\color[HTML]{FE0000} 92.78}     & {\color[HTML]{FE0000} 93.50}     & {\color[HTML]{FE0000} 94.40}     & {\color[HTML]{FE0000} 94.12}     & {\color[HTML]{FE0000} 92.55}     & {\color[HTML]{FE0000} 92.95}     & {\color[HTML]{FE0000} 92.01}     & {\color[HTML]{FE0000} 94.26}     & {\color[HTML]{FE0000} 90.86}     & {\color[HTML]{FE0000} 91.51}                      & \multicolumn{1}{c|}{{\color[HTML]{FE0000} 85.18}}       & \multicolumn{1}{c|}{\multirow{-6}{*}{\textbackslash{}}} & \multicolumn{1}{c|}{\multirow{-6}{*}{\textbackslash{}}} & \multirow{-6}{*}{\textbackslash{}} \\ \hline
\end{tabular}}
\end{table}


In order to quantitatively analyze the temporal consistency for the atlas, we define the temporal consistency(TC) factor. Suppose $I_{atlas}^{LV}(t^m)(m=1,\cdots,M)$ are the lateral ventricle maps for atlas image on the $m^{th}$ time point and $I_{atlas}^{LV}(t^m\rightarrow t^{m'})(m'=m\pm i, i = 1,2)$ are the tissue label maps warped from $t^{m'}$ to $t^{m}$. Then the temporal consistency(TC) factor can be calculated as:
\begin{equation}
TC(t^{m}) = \frac{1}{|m'|}\sum\limits_{m'}DICE\left(I_{atlas}^{LV}\left(t^{m'}\right), I_{atlas}^{LV}\left(t^{m}\rightarrow t^{m'}\right)\right)
\end{equation}
Thus, TCs of lateral ventricle reflect the temporal consistency of the tissue maps. Higher values indicate relatively better temporally consistent results. As shown in Table~\ref{tab1}, our refined atlases correct the inconsistency time point and apparently improve the temporal consistency.

\section{Conclusion}
In this paper, we propose a multi-stage implicit neural representation framework to train and construct a longitudinally consistent 4D fetal brain atlas. Experimental results demonstrate that the denoised 4D fetal brain atlases achieve better time consistency and good brain structure representation both qualitatively and quantitatively. In addition, our framework can be extensively applied to other atlases and the continuous longitudinal fetus brain atlases we constructed can also be extensively applied to other tasks, such as constructing finer 4D atlases in both spatial or temporal resolution.

%
%
%
\newpage
\bibliographystyle{splncs04}
\bibliography{ref}
\end{document}